\documentclass[twocolumn,prl, amsmath,amssymb]{revtex4}

 \usepackage{bm, color}

\usepackage{graphicx}
\usepackage{epstopdf}

\newcommand{\be}{\begin{equation}}
\newcommand{\ee}{\end{equation}}

\newcommand{\la}{\langle}
\newcommand{\ra}{\rangle}

\newcommand{\eps}{\epsilon}

\begin{document}

\title{Effect of half-quantum vortices on magnetoresistance of perforated  superconducting films.}

\author{Victor Vakaryuk}
\email[]{vakaryuk@anl.gov}
\author{Valerii Vinokur}

\affiliation{Materials Science Division, Argonne National Laboratory, Argonne, Illinois 60439, USA}


\begin{abstract}
Recent  cantilever magnetometry measurements of annular
micron-size samples of $\rm Sr_2RuO_4$ have revealed 
evidence for the existence of half-quantum vortices (HQVs) in
this material \cite{Jang:2011}. 
We propose to look for HQVs in transport measurements 
and calculate magnetoresistance of a perforated superconducting film close to the transition temperature in the presence of HQVs.
We analyze the dependence of magnetoresistance on the thermodynamic stability of
HQVs which according to \cite{Jang:2011} can be varied by the application of an in-plane magnetic field and point out features which may help to identify them.
\end{abstract}


\maketitle

\emph{Introduction.}  Half-quantum vortex (HQV) is a topological defect in a two-component superfluid system which is defined as a state in which one of the components has an extra quantum of vorticity relative to the other. $^3$He-A, where the role of the two components is played by the spin projections, has been a long standing candidate  in search for HQVs \cite{Volovik, Cross} however up to date the convincing evidence for the existence of HQVs in that system is lacking \cite{HQVsinHe3:2008}. More recently cantilever magnetometry measurements in micron-size $\rm Sr_2RuO_4$ (SRO) samples, a candidate spin-triplet superconductor with the pairing symmetry which is believed to be analogous to that of $^3$He-A, have revealed evidence for the existence of HQVs in this material \cite{Jang:2011}.

	\begin{figure}
	\includegraphics[scale=0.5]{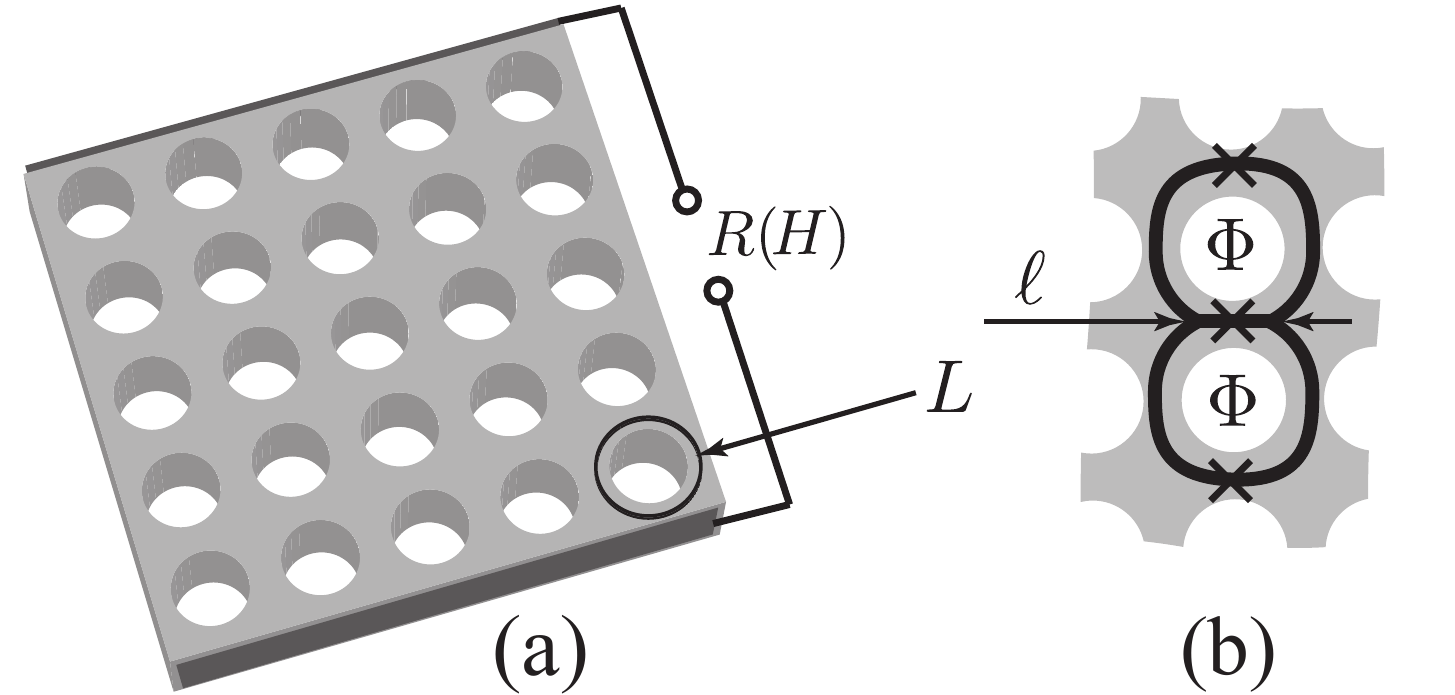}
	\caption{\label{fig:patterned film} (a) Proposed setup for detection of HQVs. Relevant dimensions are discussed in the text. (b) Elementary block model used for the calculation of magnetoresistance. See text for details.
	\vspace{-10pt}
	}
\end{figure}

Recently revived interest in HQVs stems from their potential to host Majorana modes which in turn can provide a platform for topological quantum computing\! \cite{Nayak:2008}. From a practical point of view the ability of SRO to support HQVs, if established, would provide a definitive proof for the unconventional nature of its pairing state. It is thus desirable, for both reasons, to have more experiments which could test the existence of HQVs in SRO. One possibility is to use scanning flux imaging employing either Hall \cite{Grigorenko:2003, Harada:1996} or SQUID \cite{Koshnick:2007, SQUIDlimits:2010} probes. However these  techniques are technologically quite demanding and time consuming and not as readily available in experimental labs as e.g.~transport probes.

 While most of the experimental work on SRO up to date has focused on bulk samples, recent technological advances made it possible to grow c-axis oriented epitaxial films of SRO with thickness comparable to the zero-temperature coherence length and a reasonable transition temperature \cite{SROfilms:2010}. It is known that in superconducting films with nontrivial topology such as perforated  films  visualized  as a collection of regularly spaced openings in an otherwise uniform medium (``swiss cheese'' geometry, see Fig.~\ref{fig:patterned film}a), variation of external magnetic field  can produce resistance oscillations in the vicinity of the transition temperature \cite{Tinkham:1983, Pannetier:1984, Baturina:2003, Baturina:2011}. It is often a case that the periodicity of the most pronounced resistance oscillations corresponds to a flux quantum of the applied field through the smallest opening suggesting that such oscillations are caused by vortex entrances.   Here we consider qualitative features of magnetoresistance (MR) of perforated superconducting films in the presence of HQVs and point out to the signatures of HQVs in MR curves.

Use of thin films in search for HQVs has a distinct advantage that screening effects which in the bulk hinder the thermodynamic stability of HQV relative to that of a full vortex will be substantially reduced. 
The perforated geometry which can be viewed as a collection of many small openings also offers advantage of relatively straightforward contact attachment which can be quite challenging for a small  isolated SRO annulus. 
The downside of a resistive measurement is that it should be done at temperatures close to $T_c$ where the phase stiffness of a spin current approaches that of a charge current \cite{Leggett:1975} which makes HQVs less stable. However one of the key findings of ref.~\cite{Jang:2011} is that the stability region of features attributed to HQVs can be expanded by applying in-plane magnetic field.  Coupling to the in-plane field can provide a mechanism for  stabilization of HQVs and hence can be used to both induce and identify their presence in the MR of thin films.

\emph{Magnetoresistance of  perforated superconducting films.}  Experimentally MR of thin perforated superconducting films shows an oscillating behavior  with a monotonically decaying envelop caused by a global suppression of the superfluid density; the most pronounced  oscillation period corresponds to a flux quantum of the applied field through a smallest opening \cite{Pannetier:1984, Tinkham:1983, Bezryadin:1995, Sochnikov:2010, Baturina:2011}. In high-quality arrays  a fine structure such as less pronounced additional MR minima in each period has been observed and correspond to  a flux quantum shared by several lattice cells. Our task is to show how to distinguish between these regular features and the ones induced by HQVs.

A necessary condition for a film to produce observable MR oscillations is that it should be in a resistive state. One of the factors which controls the temperature range of the resistive state  is the thickness of the film. While we do not see fundamental limitations on the film thickness apart from those set by obvious superconducting length scales, use of thinner films will broad the resistive transition thus facilitating the observation of MR oscillations. We should also point out that well defined MR features can be observed only in periodic arrays. After analyzing available data from different groups authors of ref.~\cite{Baturina:2011} concluded that one of the conditions for the observation of  MR oscillations  is that the width of the constriction separating neighboring openings should be less than $4\, \xi(T)$. However from the point of view of thermodynamic stability of HQVs for  films with thickness $w \! \lesssim \! \lambda$ it is beneficial to reduce the constriction width below the Pearl penetration depth $\lambda^2 \! /w$  which for SRO where $\lambda/\xi \! \approx \! 2.5$ can be larger  or smaller than $4\, \xi(T)$ depending on the film thickness. Thus a priori to improve the possibility of seeing HQVs in perforated SRO films for  films with $w\! \lesssim \! \lambda(T)$ openings can be spaced further apart (but not more than $4 \xi(T)$ for MR) while for thicker films openings should be spaced as close as possible.

\emph{Theoretical model.}  We now proceed to the calculation of magnetoresistance (MR)  of a perforated superconducting film close to $T_c$. Existing approaches to this problem fall into two groups. In the first group one considers a periodic array of openings and calculates  variation of $T_c$ induced by the applied field $H$; it is then argued that the MR will trace the dependence $T_c(H)$. Such approach has been adopted e.g.~for the proximity induced Josephson link arrays  \cite{Tinkham:1983},  for a regular square network of superconducting aluminum \cite{Pannetier:1984} and for thin amorphous TiN films which show superconductor-insulator transition \cite{Baturina:2011}.

Approaches in the second group are based on the observation that the most pronounced features in the MR  are correlated with a quantized value of the flux though a single opening so that instead of considering the resistance of the whole lattice one can focus on determining  the resistance of a much smaller  (``elementary'') block which usually consists of just one \cite{Sochnikov:2010} or few  \cite{Tinkham:1983} lattice cells. Given a particular block structure it is then necessary to establish a dissipative mechanism responsible for a finite resistance. Close to $T_c$ where $\xi$ becomes comparable to the width of constriction between the openings the mechanism  responsible for the oscillatory part of MR is provided by phase slips  generated by thermally activated vortex jumps in and out of openings.  It is then argued that such phase slips are described by the Ambegaokar-Halperin (AH) model of a thermally activated resistance of a heavily damped Josephson junction \cite{AH:1969}.

AH considered RCSJ model of a Josephson junction supplemented by a thermal voltage noise. They showed that this problem is equivalent to the problem of Brownian motion of a particle in the ``washboard'' potential and, in the case of large damping, obtained $I$-$V$ curves of the junction. In the limit of  vanishingly small currents the ratio of the junction's resistivity to its normal state value $R_n$
is given by
\vspace{-4pt}
\be
	R/R_n= I_0^{-2} (\Delta E /2 k_{\rm B} T)
	\label{def of R}
\vspace{-3pt}
\ee
where $I_0$ is a modified Bessel function and $\Delta E$ is the activation barrier.

The AH model has been successfully applied to a variety of physical situations seemingly beyond its original scope. In particular, assuming that vortex passages can be treated as phase slips and choosing appropriate activation barrier Tinkham has applied eqn.~(\ref{def of R}) to explain the width and the shape of the irreversibility line in some cuprate superconductors \cite{Tinkham:1988}. More recently Sochnikov et al.~\cite{Sochnikov:2010} have successfully  modeled magnetoresistance of nanopatterned superconducting LSCO arrays using eqn.~(\ref{def of R}) with the activation barrier determined by the energy level spacings between vortex states of array's openings. We will adopt a similar approach and use the AH model to point out  features in the oscillatory part of MR which are brought about by the presence of HQVs.

One can argue that application of the AH model to multiply connected geometries in the magnetic field requires further justification since eqn.~(\ref{def of R}) was derived under the assumption of  vanishingly smalls currents  while in the magnetic field, even in the absence of transport current, there generally exists non-zero screening current. To justify the use of eqn.~(\ref{def of R}) for that case we consider a pair of Josephson junctions threaded by flux $\Phi$ (dc-squid) with each junction described by the RCSJ model with thermal voltage noise. By writing down a set of coupled equations  for phase drops $\theta_1$ and $\theta_2$ of the junctions it can be shown that under the assumption of  \emph{negligible flux noise} this model describes  a single junction with phase drop $(\theta_2 - \theta_1)/2$ and a flux-dependent activation barrier and is thus equivalent to the AH model.

Thus we have reduced the problem of calculating of MR of a perforated superconducting film to a calculation of a flux-dependent activation barrier of an elementary block. Further simplification is reached by noticing that provided the ratio of the Josephson energy to the superfluid density is not very small the shape of the activation barrier of a dc-squid is well represented by the spacing between energy states with different vorticity of a junction-free loop which can be checked through their graphical comparison. This observation allows us to replace a dc-squid in the elementary block with a junction-free loop with uniform phase winding. 

For the reasons which will become clear later we consider an elementary block which is composed of two lattice cells and is represented by two overlapping loops with circumference $L$ each, the overlap length $\ell$ and cross-section $A$  as on Fig.~\ref{fig:patterned film}b. In the limit $T \to T_c$  the radial variations of the order parameter can be neglected and each loop which can support HQVs or, more precisely,  half-integer fluxoids  should be characterized by two winding numbers which we denote as $n_{si}$ and $n_{spi}$, $i=1,2$ \cite{Chung:2007, Vakaryuk:2009}. In this notation full vortices are described by $n_{si} \in \text{\small integer}$ and $n_{spi} =0$ while for HQVs $n_{si} \in \text{\small half-integer}$ and $n_{spi} =\pm 1/2$.

We are now in the position to consider energy levels of an elementary block which, according to the discussion above, can be used to compute activation energy barriers needed for the calculation of the oscillating part of MR. We notice that close to $T_c$ the screening is weak so that instead of Gibbs potential one can use free energy; the reduced screening also allows us to ignore the inductive coupling between the loops. Denoting winding numbers of each loop $n_{si}$, $n_{spi}$ and introducing $n_{s\pm} \equiv (n_{s1}\pm n_{s2})/2$ and,  analogously, $n_{sp\pm} \equiv (n_{sp1}\pm n_{sp2})/2$ the free energy of the two-cell elementary block with each loop pierced by applied flux $\Phi$ is given by
\vspace{-3pt}
\be
\begin{split}
	F = &
	\eps_0 \!
	\left[
	\left(  n_{s+}\!  - \!  \Phi/\Phi_0 \right)^2 \! +\! \gamma  n_{s-}^2
	\! + \! 
	\frac{\rho_{sp}}{\rho_s}
	\left(
	 n_{sp+}^2   \! +\!  \gamma n_{sp-}^2
	\right) 
	\right]
	\\
	 & +  
	\Delta \eps_{\textsc{hqv}},
	\label{F two loops}
\end{split}
\vspace{-3pt}
\ee
where $\eps_0 \equiv {A \Phi_0^2 }/{(4\pi \lambda^2(L-\ell) )}$ and $\gamma \equiv (L-\ell)/ (L+\ell) <1$ is a numerical factor determined by the geometry of the elementary block. 

	\begin{figure}
	\center
	\includegraphics[scale=0.5]{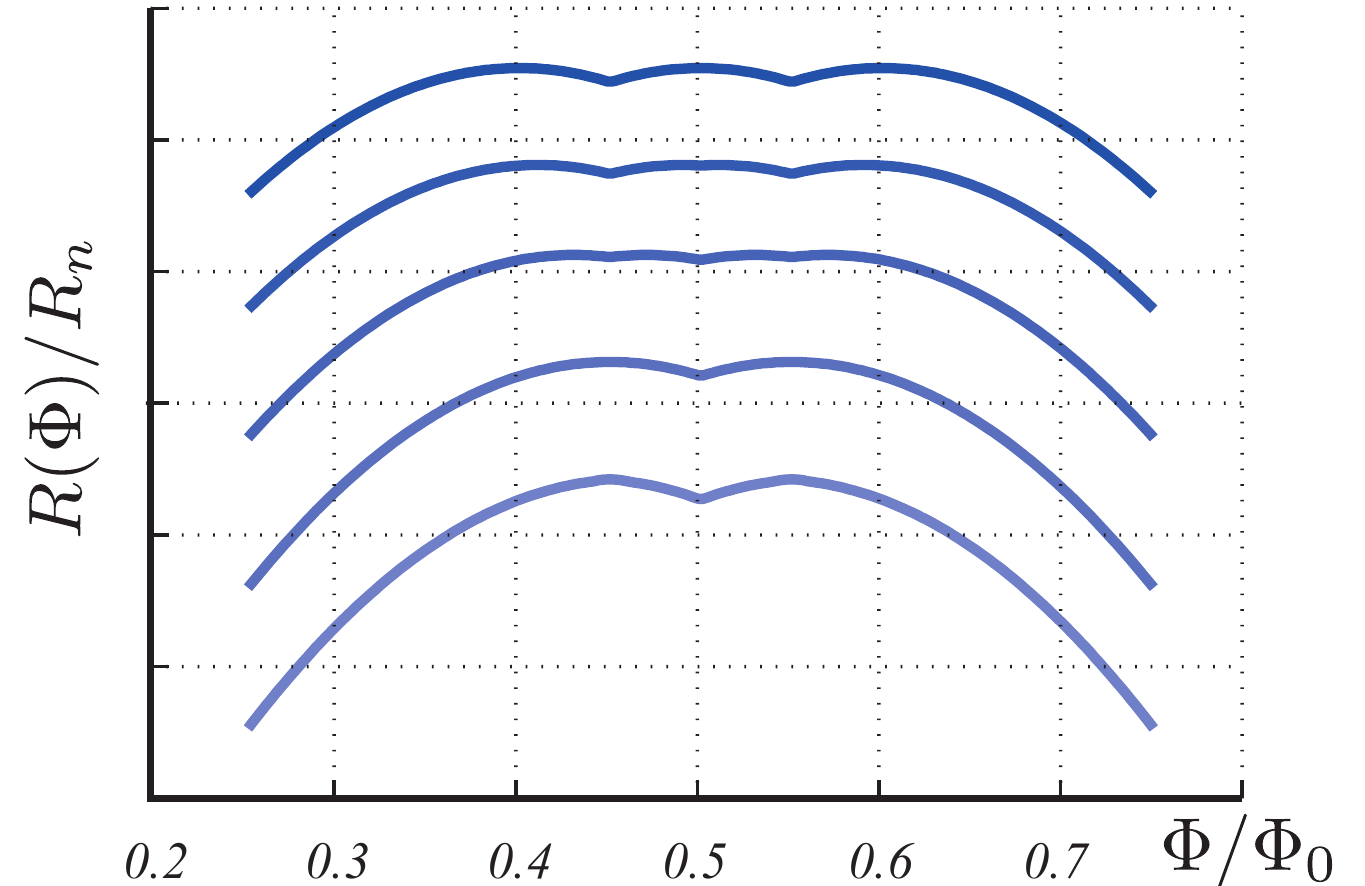}
	\caption{\label{fig:resistance} Magnetoresistance of a perforated film around $\Phi/\Phi_0=1/2$ in  the presence of HQVs for different values of the in-plane field. The curves are offset for clarity and correspond to the in-plane field $H_x$ determined by, from the bottom to the top, $h(H_x)/\eps_0 = 0.04$, 0.05, 0.07, 0.09 and 0.10. All curves are evaluated at $\gamma=4/5$ and $\eps_{ld}/\eps_0 = 0.05$ for which $H_{x0}$ is given by $h(H_{x0})/\eps_0 = 0.05$.
	\vspace{-10pt}
	}
\end{figure}

The free energy specified by eqn.~(\ref{F two loops}) contains two contributions. The first contribution proportional to the term in the brackets -- the kinematic term -- is the kinetic energy of charge and spin currents  which depends on $\rho_s$ and $\rho_{sp}$ -- superfluid and spin superfluid densities, which characterize stiffness of charge and spin currents and satisfy the inequality $\rho_{sp}/\rho_s <1$ \cite{Leggett:1975}. If the screening is ignored the latter condition implies that the kinematic term alone can provide the stability of HQV \cite{Chung:2007, Vakaryuk:2009}.  However upon approaching the transition temperature $\rho_{sp}/\rho_s \to 1$ so that the kinematic contribution by itself cannot guarantee the thermodynamic stability of HQV.

The second contribution in eqn.~(\ref{F two loops})  denoted as $\Delta \eps_{\textsc{hqv}}$ contains terms which are explicitly flux-independent e.g.~those coming from spin-orbit interactions. Recent cantilever magnetometry measurements  of micron-size SRO annuli \cite{Jang:2011}, if interpreted in terms of HQVs, imply  that the magnitude and the sign of this term can be controlled by the in-plane field -- the magnetic field directed along the ab-plane which in our geometry coincides with the plane of the film and hence does not contribute to the flux $\Phi$. While the origin of such in-plane coupling is still unclear, it may be attributed to the effect of kinematic spin polarization  according to which the HQV state posses additional spin polarization not present in the full vortex state \cite{Vakaryuk:2009}. Coupling of the kinematic spin polarization to the in-plane field requires that the pairing symmetry of SRO is such that the axis of ``easy'' spin polarization -- so-called equal-spin-pairing axis -- has a non-zero in-plane component. 

Generalizing the in-plane field coupling model proposed in \cite{Jang:2011} for a two-loop elementary block $\Delta \eps_{\textsc{hqv}}$ can be written as
\vspace{-3pt}
\be
	\Delta \eps_{\textsc{hqv}}
	=
	\eps_{ld} (| n_{sp1}| + | n_{sp1}|)- h(H_x)( n_{sp1} +n_{sp2})
	\label{delta hqv}
\vspace{-3pt}
\ee
where $\eps_{ld}>0$  is a term which originates from the spin-orbit energy difference between the half-quantum and full vortex states and $h(H_x)$ is a monotonic function of the in-plane field $H_x$; notice that the coefficient in front of $h(H_x)$ is proportional to the sum of spin currents in each loop in accordance with the kinematic spin polarization model \cite{Vakaryuk:2009}. As observed in \cite{Jang:2011} for micron-size SRO annuli, contribution $\Delta \eps_{\textsc{hqv}}$ which is non-zero only in the HQV states, becomes negative as the in-plane field reaches a sample-dependent value $H_{x0}$ which lies  in the range 10\,G\,-200\,G thus inducing the thermodynamic stability of HQV. In our model $H_{x0}$ is defined by the condition $h(H_{x0})=\eps_{ld}$.

The full set of energy levels described by eqn.~(\ref{F two loops}) is periodic in flux and is degenerate with respect to the interchange of the two cells in the elementary block; we will ignore this degeneracy since in real-life situations it will be removed by the presence of the sample's boundary. Complexity of the energy levels grows substantially  with their energy and depends on the value of parameters entering eqn.~(\ref{F two loops}). In the absence of the in-plane field the lowest energy level consists of three full vortex states. Denoting each state by a set of four winding numbers $\{n_{s1}, n_{sp1}, n_{s1}, n_{sp2}\}$ and setting $\rho_{sp}/\rho_s \!=\!1$ as appropriate for temperatures close to $T_c$, the lowest energy level is realized by $\{0,0,0,0\}$ for $\Phi/\Phi_0 \! \in \! [0, (1+\gamma)/4]$, $\{1,0,0,0\}$ for $\Phi/\Phi_0\! \in \! [(1+\gamma)/4, (3-\gamma)/4]$ and $\{1,0,1,0\}$ for $\Phi/\Phi_0 \! \in \!  [(3-\gamma)/4, 1]$ which corresponds to no vortices, one vortex and two full vortices in the elementary block. 

As  the in-plane field reaches value $H_{x0}$ defined above the lowest energy level acquires  contributions from states $\{1/2, 1/2, 0,0\}$ and $\{1/2,1/2, 1,0\}$ which correspond to an HQV in one of the cells. Notice that states with HQVs in each cell will either not couple to the in-plane field (state $\{1/2,1/2,1/2,-1/2\}$) or will contribute to the ground state level only starting from a higher value $H_{x1}$ of the in-plane field  (state $\{1/2,1/2,1/2,1/2\}$) defined by $h(H_{x1}) = \eps_{ld} + \eps_0 (1- \gamma)/4$.

\emph{Results.} The oscillating part of magnetoresistance is calculated using eqn.~(\ref{def of R}) choosing the activation barrier as a spacing between two lowest energy levels of model (\ref{F two loops});
the result  is shown on Fig.~\ref{fig:resistance} for different values of the in-plane field. Zooming in the neighborhood of $\Phi/\Phi_0 = 1/2$ we see that for in-plane fields $H_x < H_{x0}$ magnetoresistance has a dip  at $\Phi/\Phi_0 =1/2$; this dip is produced by a configuration with one full vortex per  two-cell elementary block and can be observed in conventional superconducting films, see e.g.~\cite{Tinkham:1983}. Upon increasing the in-plane field above $H_{x0}$ states with one HQV per block produce  satellite dips located symmetrically  around $\Phi/\Phi_0 = 1/2$ while the central minimum becomes shallow and eventually disappear. Unlike the minimum at $\Phi/\Phi_0 = 1/2$ such behavior is characteristic only to the films which are able to support HQVs. For in-plane fields just above $H_{x0}$  location of the satellite HQV minima $\Phi_{\text{\tiny HQV}}$ is field-independent and is given by $|\Phi_{\textsc{\tiny HQV}}/\Phi_0 - 1/2| = (1-\gamma)/4$.

The presence of higher energy levels can be accounted for by using a procedure suggested by Sochnikov et al.~\cite{Sochnikov:2010}.
This procedure consists of replacing $\Delta E$ calculated on the two lowest energy  levels with the Gibbs average $\la \Delta E \ra$ calculated using full spectrum described by (\ref{F two loops}). We have checked that such  procedure while able to smooth out magnetoresistance features does not change their evolution with the in-plane field as described above.

It should be noted that conventional vortex configurations can also produce magnetoresistance minima which are located at the rational values of the applied flux. These features, which are usually less pronounced,  correspond to full vortex states of larger elementary blocks and can only be observed in high-quality films \cite{Pannetier:1984, Baturina:2011}. However, unlike minima  produced by HQVs, minima due to conventional vortices will not evolve with the in-plane field in the way described above. Hence the in-plane field evolution of HQVs as observed in \cite{Jang:2011} can be used to identify their magnetoresistance signatures.

\emph{Conclusion.}  We have proposed to search for half-quantum vortices in transport properties of perforated superconducting films in the presence of magnetic field. In particular, we have calculated magnetoresistance of a perforated superconducting film assuming that the film's material is able to support half-quantum vortices as has been recently experimentally suggested for $\rm Sr_2 Ru O_4$ in cantilever magnetometry measurements \cite{Jang:2011}. We have shown that the oscillating part of the magnetoresistance should have  well defined signatures of HQVs which, given the findings of ref.~\cite{Jang:2011} can be controlled by applying the in-plane magnetic field.

Authors would like to thank Tatiana Baturina for useful discussions. The first author would like to thank David Ferguson for pointing out ref.~\cite{SROfilms:2010} and Yoshifumi Yamaoka and Yoshi Maeno for useful comments. This work was supported by the Department of Energy under the Contract No.~DE-AC02-06CH11357.


\end{document}